\documentclass[runningheads,a4paper]{llncs}

\usepackage{amssymb}
\usepackage{amsmath}
\usepackage{booktabs}
\usepackage{multirow}
\usepackage{graphicx}
\usepackage{url}
\usepackage[colorlinks,bookmarks=false]{hyperref}
\newcommand{\keywords}[1]{\par\addvspace\baselineskip
\noindent\keywordname\enspace\ignorespaces#1}

\begin{document}
\sloppy

\mainmatter

\title{Hybrid Context Forests and Bilateral Viterbi Matching for 3D Rib Centerline Tracing in CT}

\titlerunning{Hybrid Context Forests for 3D Rib Centerline Tracing in CT}

\author{Quan Wang}
\authorrunning{Q. Wang}

\institute{}

\toctitle{Hybrid Context Forests and Bilateral Viterbi Matching for 3D Rib Centerline Tracing in CT}
\tocauthor{Quan Wang}
\maketitle

\begin{abstract}
Automatic extraction and anatomical numbering (R01--R12, L01--L12) of all 24 rib centerlines in thoracic CT is essential for unfolded rib visualization, fracture detection, and surgical navigation, but remains challenging due to adjacent-bone confluence, calcified costal cartilage bridges, kyphotic pose variation, and partial visibility of terminal ribs. We present ribline, an open-source three-stage framework combining: (1) an automatically detected lung-apex and vertebral-midline coordinate frame with a kyphosis-aware azimuthal angle for point ordering along each rib; (2) a hybrid voxel detector fusing an anatomically anchored Semantic Context Forest (SCF) with a lightweight 3D U-Net; and (3) a global stage that decomposes the probability skeleton into candidate curves, assigns rib indices via a bilateral Viterbi dynamic program coupled across both hemithoraces, and refines curves via template recovery and normal-plane centering (RibSnake). On the Siemens Corporate Research 2 mm thoracic CT benchmark (47 training and 12 evaluation scans), our pipeline achieves a labeled Chamfer distance of 2.05 mm, labeled clDice of 0.886 (3 mm) and 0.945 (4 mm), 97.2\% labeling accuracy (278/286 ribs), and zero false-positive curves on the evaluation split, and 3.64 mm (median 2.39 mm) with 91.8\% labeling accuracy across all 59 annotated scans (1,402 ribs) in 5-fold cross-validation.

\textbf{Code:} \url{https://github.com/wq2012/ribline}\\
\textbf{Pretrained Models:} \url{https://huggingface.co/collections/wq2012/ribline}

\end{abstract}

\section{Introduction}
\label{sec:intro}

The human rib cage comprises twelve pairs of curved, tubular bones (\texttt{R01}--\texttt{R12}, \texttt{L01}--\texttt{L12}). Accurate 3D extraction and individual numbering of rib centerlines in CT enables unfolded rib cage visualization via curved planar reformation for trauma screening \cite{ringl2015,ribfrac2020}, provides an anatomical coordinate system for localizing thoracic lesions \cite{staal2007}, supports radiotherapy and spinal instrumentation planning \cite{klinder2007}, and is the topological backbone of automated rib-fracture detection \cite{ribseg2024}.

Despite the high attenuation of cortical bone, fully automatic 3D rib centerline tracing and numbering is hard for four reasons:
(1) \emph{adjacent bone confluence}: the first rib articulates with the manubrium and lies next to the clavicle and scapula, so skeletonization leaks into the shoulder girdle;
(2) \emph{intercostal ossification}: calcified costal cartilage and partial-volume averaging create bridges between ribs 5--10, producing false junctions and hairpin U-turns;
(3) \emph{terminal-rib variability}: the floating twelfth ribs vary widely in length and are congenitally absent in a few percent of patients, so naive top-down or bottom-up counting suffers index shifts whenever a terminal rib is missed; and
(4) \emph{thoracic kyphosis}: in the upper thorax the rib cage curves anteriorly and inferiorly while the spine arches posteriorly, so a polar angle around the torso centroid is not a reliable ordering parameter along the posterior rib neck.

Early methods relied on seeded region growing or ridge tracking \cite{aykac2003,staal2007,klinder2007,ramakrishnan2011}. Wu \textit{et al.} \cite{wu2012learning} introduced learning-based deformable template matching with Marginal Space Learning \cite{zheng2008} on the SCR dataset used here, reporting $89.5$--$95.9\,\%$ labeling accuracy under their own protocol. Auto-context \cite{tu2010} and Semantic Context Forests (SCF) \cite{wang2013semantic,wang2014thesis} showed that combining multi-scale differential features with long-range random-shift intensity/probability differences gives discriminative voxel classifiers for thin structures. Deep 3D networks \cite{lenga2018,ribfrac2020} and point-cloud networks such as RibSeg~v2 \cite{ribseg2024} achieve strong segmentation, but on small training cohorts standalone CNNs still under-segment the small first and floating twelfth ribs and need post-hoc heuristics for numbering.

\begin{figure}[t]
  \centering
  \includegraphics[width=\textwidth]{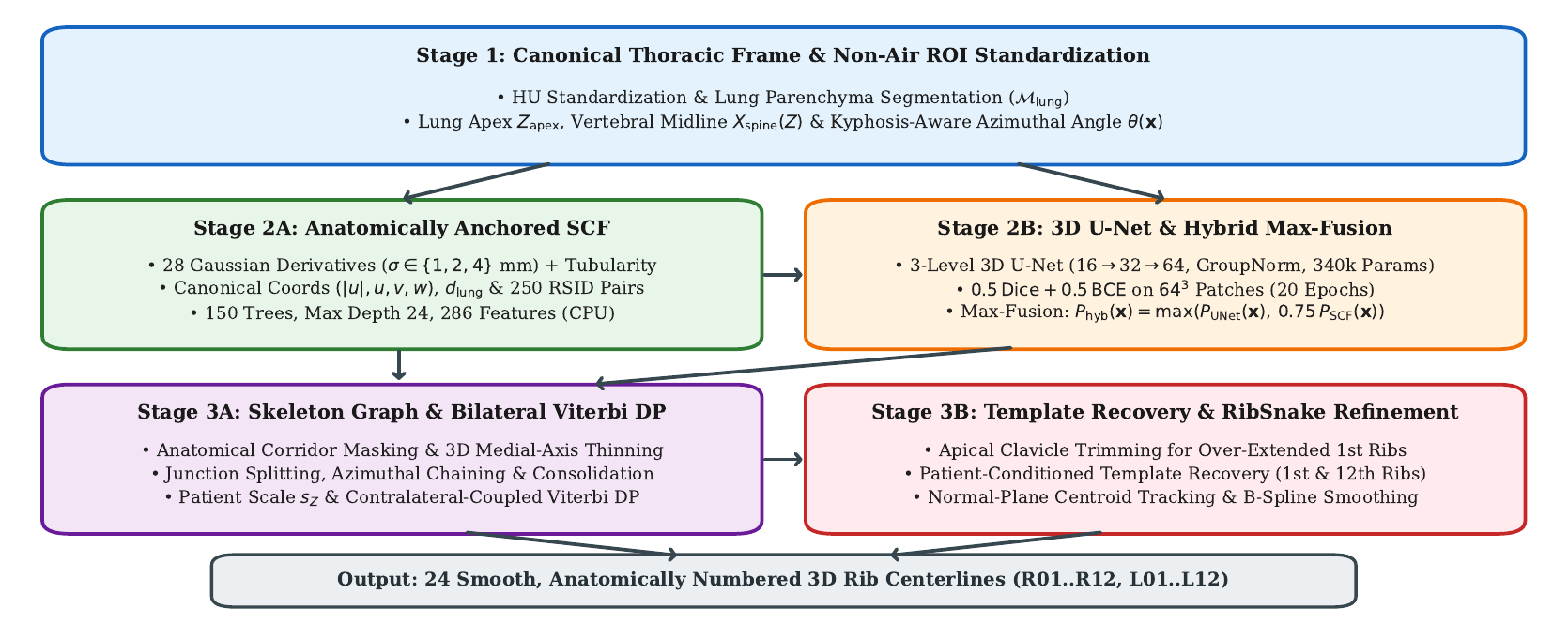}
  \caption{Architecture of the proposed three-stage \texttt{ribline} framework: Stage~1 standardizes Hounsfield units and establishes the lung-apex / vertebral-midline coordinate frame; Stage~2 fuses an anatomically anchored Semantic Context Forest (SCF) with a lightweight 3D U-Net; Stage~3 extracts the skeleton graph, solves contralateral-coupled bilateral Viterbi labeling, and refines centerlines via template recovery and \texttt{RibSnake} normal-plane centering.}
  \label{fig:pipeline}
\end{figure}

\textbf{Contributions.} We describe and release (Fig.~\ref{fig:pipeline}) a three-stage pipeline whose methods section below is written to match the released code line by line:
\begin{enumerate}
  \item a \textbf{lung-apex / vertebral-midline thoracic frame} and a \textbf{kyphosis-aware azimuthal angle} $\theta(\mathbf{x})$ whose focus moves posteriorly towards the apex; on the 1,402 annotated ribs of the 59 valid SCR scans $\theta$ increases from the posterior to the anterior end of \emph{every} rib and is strictly monotonic on $78.6\,\%$ of ribs (vs.\ $57.0\,\%$ for a torso-center angle; Sec.~\ref{sec:frame});
  \item a \textbf{complementary hybrid detector}: an anatomically anchored SCF (better on first ribs) max-fused with a plain 3D U-Net (better on intermediate ribs);
  \item a \textbf{patient-scaled bilateral Viterbi labeler} with contralateral coupling and outlier tolerance, plus \textbf{\texttt{RibSnake}} (clavicle trimming, terminal-rib template recovery, normal-plane centering);
  \item a \textbf{rigorous evaluation protocol} (symmetric labeling criterion, missed-rib penalty, precision, bootstrap and Wilson confidence intervals, paired tests), a transparent treatment of one annotation error in the public ground truth, and a cross-validation with frozen Stage-3 constants that quantifies how much the development protocol favored the 12 evaluation scans.
\end{enumerate}

\section{Canonical Thoracic Frame and Statistical Rib Atlas}
\label{sec:frame}

\begin{figure}[t]
  \centering
  \includegraphics[width=0.96\textwidth]{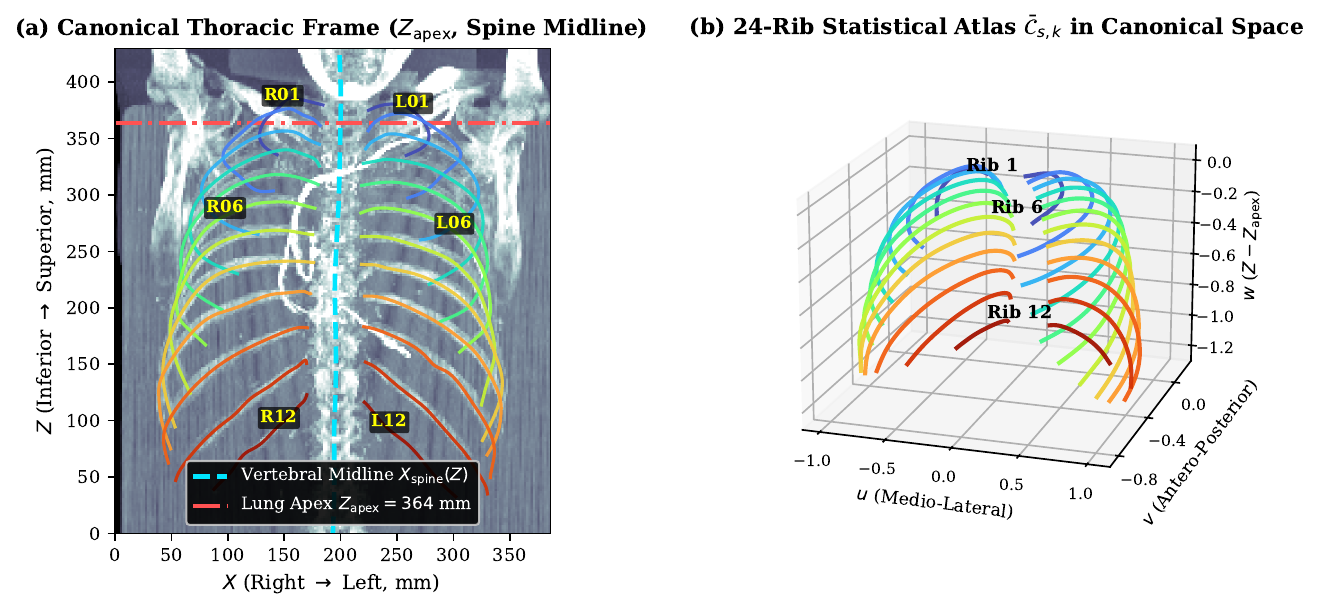}
  \caption{Canonical thoracic reference frame and statistical shape prior. (a) Coronal CT MIP with automatically detected lung apex $Z_{\mathrm{apex}}$, vertebral midline trajectory $X_{\mathrm{spine}}(Z)$, and ground-truth rib centerlines. (b) The 24-rib canonical statistical atlas $\bar{\mathcal{C}}_{s,k}$ fitted on the 47 training scans.}
  \label{fig:frame_atlas}
\end{figure}

\subsection{Hounsfield Standardization and Thoracic Landmark Detection}
The SCR archive mixes signed Hounsfield units (air $\approx -1000$) and unsigned shifted values (air $\approx 0$). If the minimum intensity of a volume is $\ge -100$, we subtract $1024$; otherwise the volume is used as is (\texttt{standardize\_hu}).

All landmarks are computed on a $2\times$ down-sampled grid. The torso mask is $I > -350$\,HU with slice-wise hole filling; internal air is the filled torso minus voxels with $I \ge -380$\,HU, morphologically opened and reduced to the (at most four) largest components with more than $800$ sub-voxels, which form the lung mask $\mathcal{M}_{\mathrm{lung}}$. The \textbf{lung apex} $Z_{\mathrm{apex}}$ is the larger of the per-side $99.5$th-percentile $Z$ coordinates of $\mathcal{M}_{\mathrm{lung}}$. The \textbf{spine} is located as the peak of a Gaussian-smoothed ($\sigma=3$ sub-voxels) axial projection of bone intensity ($\mathrm{clip}(I-120, 0, 600)$) restricted to the posterior $40\,\%$ of the antero-posterior bone extent and to $\pm 50$\,mm around the medio-lateral bone center, refined by an intensity-weighted centroid; a per-slice median of bone voxels within $\pm 32 \times \pm 36$\,mm of this seed, smoothed with a Gaussian of $\sigma = 12$ slices, yields the trajectory $(X_{\mathrm{spine}}(Z), Y_{\mathrm{spine}}(Z))$; the scalar seed values $(x_{\mathrm{spine}}, y_{\mathrm{spine}})$ are also kept. The thoracic half-width $W$ is the $96$th percentile of $|X - x_{\mathrm{spine}}|$ and $Y_{\mathrm{ant}}, Y_{\mathrm{post}}$ the $5$th/$95$th percentiles of $Y$ over bone voxels ($I > 150$\,HU) in the band $Z_{\mathrm{apex}} - 210 \le Z \le Z_{\mathrm{apex}} - 10$\,mm; $D = \max(Y_{\mathrm{post}} - Y_{\mathrm{ant}}, 100\,\text{mm})$.

Measured on the 59 valid scans (script \texttt{polar\_angle\_stats}), the posterior (spine-side) end of \texttt{R01}/\texttt{L01} lies $+11.9 \pm 9.3$\,mm above $Z_{\mathrm{apex}}$ ($n = 117$; median $13.0$\,mm, 5--95\,\% range $-2.5$ to $24.5$\,mm), and the midline error $|X_{\mathrm{spine}}(Z) - \tfrac{1}{2}(X^{\mathrm{R}}_{\mathrm{post}} + X^{\mathrm{L}}_{\mathrm{post}})|$ over 695 annotated rib pairs is $2.5 \pm 2.4$\,mm (median $1.8$\,mm). The apex is therefore a useful but not a tight anchor for the first rib; the labeler below never relies on it alone.

\subsection{Canonical Mapping and Kyphosis-Aware Azimuthal Angle}
A physical point $\mathbf{x} = (X, Y, Z)$ is mapped to canonical coordinates
\begin{equation}
  u = \frac{X - x_{\mathrm{spine}}}{W}, \qquad
  v = \frac{Y - y_{\mathrm{spine}}}{D}, \qquad
  w = \frac{Z - Z_{\mathrm{apex}}}{240\ \text{mm}},
  \label{eq:canonical}
\end{equation}
using the scalar spine seed (the per-slice trajectory is used only for the anatomical masks of Sec.~\ref{sec:stage3}). To order points along a rib from its vertebral head to its sternal end we use an azimuthal angle whose focus sits $x_c = 8$\,mm lateral of the midline and moves posteriorly towards the apex,
\begin{align}
  y_c(Z) &= Y_{\mathrm{ant}} + 0.25\,D + 0.16\,\mathrm{clip}(Z_{\mathrm{apex}} - Z,\, 0,\, 220\ \text{mm}),\\
  \theta(\mathbf{x}) &= \mathrm{atan2}\!\big(-(Y - y_c(Z)),\; |X - x_{\mathrm{spine}}| - x_c\big),
  \label{eq:polar}
\end{align}
followed by phase unwrapping along the curve. Over the 1,402 ground-truth ribs, $\theta_{\mathrm{end}} > \theta_{\mathrm{start}}$ (correct global orientation) holds for $100\,\%$ of ribs; $\theta$ is \emph{strictly} monotonic for $1{,}102$ ribs ($78.6\,\%$), the largest local reversal exceeds $0.1$\,rad on $8$ ribs ($0.6\,\%$) and $0.3$\,rad on $3$ ribs. The naive angle $\mathrm{atan2}(-(Y - y_{\mathrm{lung}}), |X - x_{\mathrm{spine}}|)$ around a fixed torso center is strictly monotonic for only $799$ ribs ($57.0\,\%$) with reversals $>0.1$\,rad on $107$ ribs ($7.6\,\%$). All downstream uses of $\theta$ (orientation, fragment chaining, angular span) are robust to small reversals.

\subsection{Non-Parametric 24-Rib Statistical Atlas}
From the 47 training scans ($1{,}116$ ribs) each rib is oriented posterior-to-anterior, resampled to $M = 80$ arc-length-uniform points, mapped by Eq.~\ref{eq:canonical} and averaged to a mean curve $\bar{\mathcal{C}}_{s,k} \in \mathbb{R}^{80 \times 3}$ per side $s$ and index $k$ (Fig.~\ref{fig:frame_atlas}b). For every $k$ we also store mean and standard deviation $(\boldsymbol{\mu}_{f,k}, \boldsymbol{\sigma}_{f,k})$ of the 5-D descriptor
$\mathbf{f}(\mathcal{C}) = [z_{\mathrm{post}},\, \bar z,\, x_{\max}/W,\, L/W,\, y_{\min}/D]$
(posterior-end height and mean height relative to $Z_{\mathrm{apex}}$, normalized lateral reach, normalized arc length, normalized most-anterior $Y$), the inter-rib spacing $(\mu_{s,k}, \sigma_{s,k})$ between the posterior ends of ribs $k$ and $k+1$, and a $15{,}000$-point cloud $\{(\tilde{\mathbf{u}}_a, \Delta z_a)\}_{a=1}^{15000}$ of canonical points $\tilde{\mathbf{u}}_a = (|u_a|, v_a, w_a)$ paired with their height offset $\Delta z_a = Z_a - z_{\mathrm{post}}$ from the rib's posterior end. For any curve or fragment $\mathcal{C}$, unweighted Euclidean $16$-nearest-neighbor regression in $(|u|, v, w)$ estimates the posterior-equivalent height from each point $\mathbf{x}_m \in \mathcal{C}$ as $z_{\mathrm{eq}}(\mathbf{x}_m) = Z_m - \frac{1}{16}\sum_{a \in \mathcal{N}_{16}(\mathbf{x}_m)} \Delta z_a$. When the posterior endpoint $\mathbf{x}_1$ lies close to the vertebral column ($|X_1 - x_{\mathrm{spine}}| < 48$\,mm and $Y_1 > y_{\mathrm{lung}} - 15$\,mm), $z_{\mathrm{post}}$ is estimated directly as the mean $Z$ of the first three posterior points (relative to $Z_{\mathrm{apex}}$); otherwise it is set to the median of $z_{\mathrm{eq}}(\mathbf{x}_m) - Z_{\mathrm{apex}}$ over the posterior-most quarter of $\mathcal{C}$, while the median of $z_{\mathrm{eq}}(\mathbf{x}_m)$ over the entire curve defines the \emph{slant-corrected posterior-equivalent height} $z_{\mathrm{sort}}(\mathcal{C})$ used to order candidates craniocaudally. We verified that re-fitting the atlas with the released \texttt{build\_atlas} script on the 47 training scans reproduces the shipped atlas exactly (maximum absolute difference $0$), and that fitting on all 59 scans does not.

\section{Hybrid Semantic Context Forest and 3D U-Net}
\label{sec:detection}

\subsection{Semantic Context Forests}
Following \cite{wang2013semantic}, voxel-wise centerline detection is cast as classification. For every voxel we extract the raw intensity and, at $\sigma \in \{1, 2, 4\}$\,mm, the Gaussian-smoothed intensity, gradient magnitude, the six second-order Gaussian derivatives and the Laplacian (28 features), plus Random Shift Intensity Differences (RSID)
\begin{equation}
  f_{\mathrm{RSID}}(\mathbf{x}; \mathbf{u}_1, \mathbf{u}_2) = I(\mathbf{x} + \mathbf{u}_1) - I(\mathbf{x} + \mathbf{u}_2),
  \label{eq:rsid}
\end{equation}
with integer voxel offsets drawn uniformly from $[-r, r]^3$. Three forests are released (scikit-learn random forests, balanced class weights, $\sqrt{d}$ features per split):
\textbf{Pass-1} ($120$ trees, unlimited depth, $228$ features): $200$ RSID pairs with $r = 20$ voxels ($\pm 40$\,mm).
\textbf{Pass-2 auto-context} ($100$ trees, $306$ features): the Pass-1 features, $28$ multi-scale features of the Pass-1 probability map $P^{(1)}$ and $50$ Random Shift Probability Differences $P^{(1)}(\mathbf{x} + \mathbf{v}_1) - P^{(1)}(\mathbf{x} + \mathbf{v}_2)$ with $r = 40$ voxels ($\pm 80$\,mm) \cite{tu2010}.
\textbf{Anatomically anchored SCF} ($150$ trees, depth $24$, $286$ features): Hounsfield-standardized input, $250$ RSID pairs with $r = 12$ voxels ($\pm 24$\,mm), a per-scale tubularity proxy built from the diagonal second derivatives, $\tau_\sigma = (-\min(0,I_{xx}) - \min(0,I_{yy}) - \min(0,I_{zz}))\, \rho/(\rho + 50)$ with $\rho = \|(I_{xx}, I_{yy}, I_{zz})\|$ (no eigen-decomposition or Frangi filter is computed), and five anatomical channels $|u|, u, v, w$ and the distance to the lung mask. Training samples $50{,}000$ voxels per scan, half on the annotated centerlines and half uniformly over the whole volume; the Pass-1 forest additionally used random rigid augmentation of the training scan. At inference the forests are evaluated only on voxels with $G_{1.2}(I) > -20$\,HU, i.e.\ everything except air; this ROI is a speed optimization and is not used during training.

\subsection{3D U-Net and Hybrid Fusion}
The second detector is a plain three-level 3D U-Net ($16 \to 32 \to 64$ channels, two $3^3$ convolutions with GroupNorm and ReLU per level, max-pooling and transposed-convolution up-sampling, no residual connections; $339{,}889$ parameters). It is trained on $64^3$ patches ($70\,\%$ centered near a centerline voxel, $30\,\%$ uniformly), input normalized by a per-volume z-score of the \emph{raw} stored intensities (no Hounsfield standardization), target = centerline voxels dilated to a tube, loss $= 0.5\,(1 - \text{soft Dice}) + 0.5\,\text{BCE}$, Adam ($10^{-3}$), batch 2, 8 patches per scan and epoch, 20 epochs, no augmentation. The released checkpoint \texttt{rib\_unet\_3d.pth} is the epoch-20 weights; it reproduces the probability maps used for all results below bit-exactly (maximum absolute difference $0$ on re-inference). Inference uses $64^3$ sliding windows with stride 48 and averaging.

\begin{figure}[t]
  \centering
  \includegraphics[width=\textwidth]{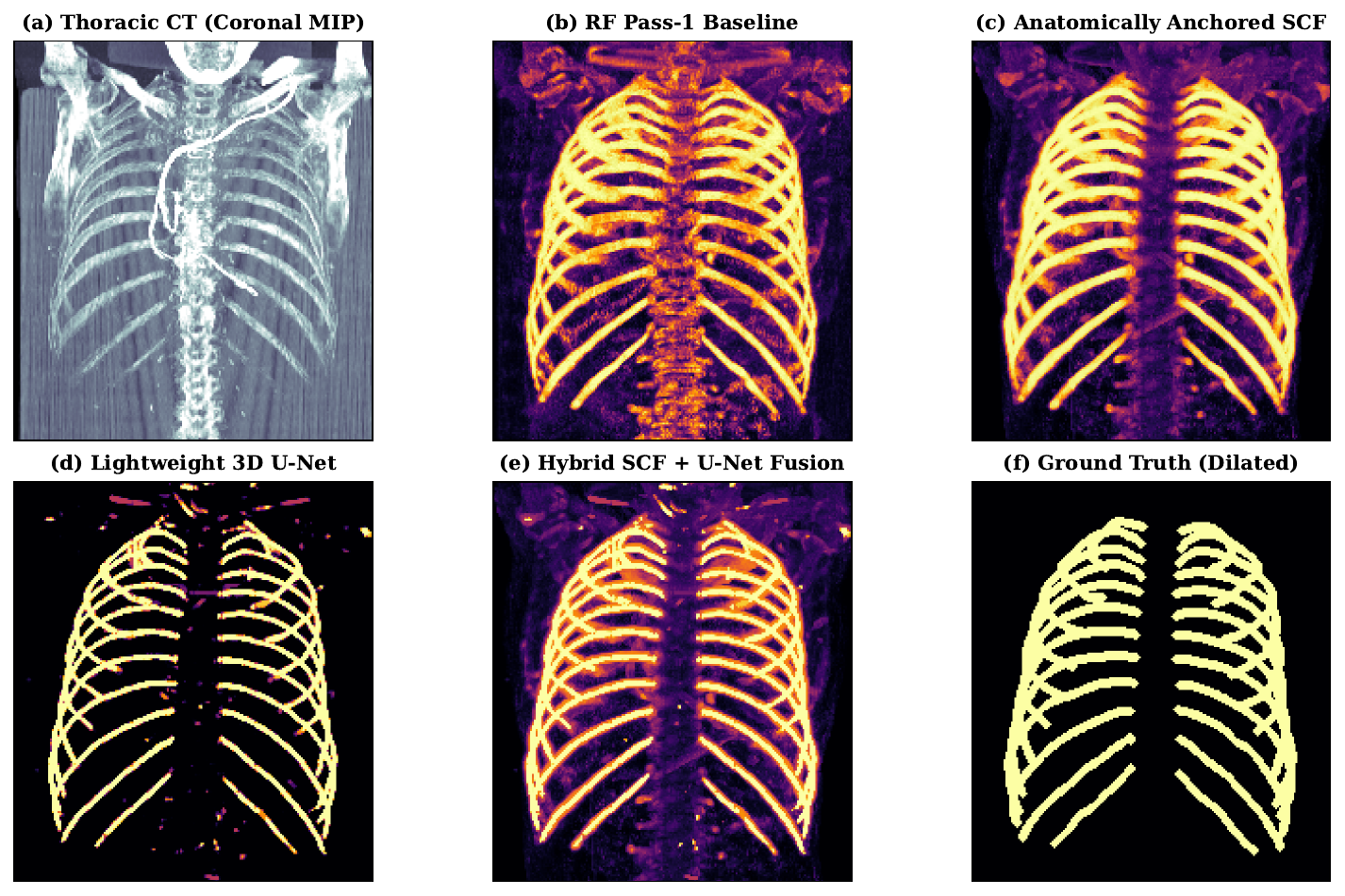}
  \caption{Coronal MIPs of centerline probability maps on evaluation scan \texttt{pet\_ct\_bc\_\_eh\_Anonymous\_0004}: (a) CT, (b) Pass-1 forest, (c) anatomically anchored SCF, (d) 3D U-Net, (e) hybrid max-fusion, (f) dilated ground truth.}
  \label{fig:prob_map}
\end{figure}

The two detectors are complementary (Fig.~\ref{fig:prob_map}): the SCF, with explicit canonical coordinates, is more sensitive on the short apical first ribs, whereas the U-Net gives cleaner shafts on intermediate ribs. We fuse the probability volumes by a voxel-wise maximum,
\begin{equation}
  P_{\mathrm{hyb}}(\mathbf{x}) = \max\!\left(P_{\mathrm{UNet}}(\mathbf{x}),\; 0.75\, P_{\mathrm{SCF}}(\mathbf{x})\right).
  \label{eq:hybrid}
\end{equation}

\section{Stage 3: Topology Matching and Deformable Refinement}
\label{sec:stage3}

\subsection{Anatomical Masking, Skeletonization and Fragment Chaining}
$P_{\mathrm{hyb}}$ is multiplied by an anatomical mask that keeps voxels within $3.5$\,mm of bone ($I > 115$\,HU or $G_{1.2}(I) > 75$\,HU), inside $[Z_{\mathrm{apex}} - 365, Z_{\mathrm{apex}} + 32]$\,mm, outside the vertebral corridor ($|X - X_{\mathrm{spine}}(Z)| < 10.5$\,mm posteriorly) and sternal corridor ($< 31$\,mm anteriorly), and, in the upper thorax ($Z > Z_{\mathrm{apex}} - 140$\,mm), closer than $26$\,mm to the lung (removing scapula and lateral clavicle). Vertical intercostal valleys ($G_{1.2}(I)$ at least $12$\,HU below both axial neighbors and $I < 135$\,HU) are suppressed. After a light Gaussian ($\sigma = 0.55$ voxels) the map is thresholded at $0.45$ and skeletonized \cite{lee1994}. The 26-connected skeleton graph is pruned of spurs shorter than $8$\,mm, junction voxels are removed to obtain elementary chains, short steep vertical bridges ($< 28$\,mm with $|\Delta Z| / L > 0.72$) are discarded, chains are split at hairpin turns of the smoothed $\theta$ profile and at jumps of the slant-corrected height larger than $16$\,mm, and collinear fragments on the same hemithorax are chained posterior-to-anterior when their gap is $< 44$\,mm, their heights agree within $12$\,mm, and the end tangents are aligned with the chord ($\cos > 0.62$; $> 0.72$ for touching fragments). Curves shorter than $22$\,mm are dropped.

\subsection{Candidate Consolidation and Patient Scale}
Per hemithorax (side by the mean $X$ relative to $x_{\mathrm{spine}}$), candidates are oriented by $\theta$, described by $(\mathbf{f}, z_{\mathrm{sort}}, \theta_{\min}, \theta_{\max}, L)$, and consolidated: fragments that are disjoint in $\theta$, within $11.5$\,mm in $z_{\mathrm{sort}}$ and $< 75$\,mm apart are bridged (greedy, closest first); upward-sloping anterior fragments ($L < 140$\,mm, start $> 30$\,mm from the midline, rising $> 3$\,mm) and short lateral fragments ($L < 85$\,mm starting $> 66$\,mm from the midline or with $\theta_{\min} > 0.25$) are removed when at least six candidates remain; and of two candidates within $8.5$\,mm in $z_{\mathrm{sort}}$ whose mean curve distance is $< 6.5$\,mm (or that are within $5.5$\,mm and very unequal in length) only the longer is kept. A patient scale $s_Z \in [0.80, 1.25]$ is the blend $0.45\, s_{\mathrm{step}} + 0.55\, s_{\mathrm{span}}$ of the median adjacent step of long ($\ge 50$\,mm) candidates divided by $21.8$\,mm and the ratio of the observed to the atlas craniocaudal span.

\subsection{Bilateral Viterbi Labeling}
For candidate $i$ and label $k \in \{1, \dots, 12\}$ we rescale the height coordinates by $s_Z$ relative to the nominal $+15$\,mm ($15/240$) first-rib apex anchor, forming $\hat{f}_{i,d} = (f_{i,d} - 15\,\text{mm})/s_Z + 15\,\text{mm}$ for $d \in \{1,2\}$ (with $\hat{f}_{i,d} = f_{i,d}$ for $d \in \{3,4,5\}$) and the scale-adjusted canonical projection $\Phi_{s_Z}(\mathbf{x}) = (u,\, v,\, \hat{w})$ with $\hat{w} = (w - 15/240)/s_Z + 15/240$. With floored standard deviations $\tilde{\sigma}_{f,k,d} = \max(\sigma_{f,k,d},\, \sigma_{\min,d})$ where $\boldsymbol{\sigma}_{\min} = (14\,\text{mm}, 14\,\text{mm}, 0.10, 0.25, 0.12)$, the raw unary cost is
\begin{equation}
  \tilde U(i, k) = \tfrac{1}{2}\sum_{d=1}^{5} w_d \Big(\tfrac{\hat f_{i,d} - \mu_{f,k,d}}{\tilde{\sigma}_{f,k,d}}\Big)^{2} + \lambda_{\mathrm{tpl}}\,\overline{d}\big(\Phi_{s_Z}(\mathcal{C}_i), \bar{\mathcal{C}}_{s,k}\big),
  \label{eq:unary}
\end{equation}
where $\lambda_{\mathrm{tpl}} = 10$ across all candidates, $\overline{d}$ is the mean one-directional nearest-neighbor distance from $40$ arc-length-resampled points of $\Phi_{s_Z}(\mathcal{C}_i)$ to the $80$-point template $\bar{\mathcal{C}}_{s,k}$, and the descriptor weights are $\mathbf{w} = (1.1, 0.7, 1.15, 1.05, 0.45)$ for full-span candidates (or $k \in \{1, 12\}$) vs.\ $\mathbf{w} = (1.2, 0.6, 0.3, 0.2, 0.25)$ for partial intermediate candidates (angular span $< 1.35$\,rad, $k \notin \{1, 12\}$). Costs are made relative, $U(i,k) = \tilde U(i,k) - \min_m \tilde U(i,m)$, with a non-negative re-added offset $\max(0,\, \min_m \tilde U(i,m) - 8)$ for short candidates ($L < 45$\,mm) so that only genuinely atypical short fragments can become clutter. Left/right candidates whose $z_{\mathrm{sort}}$ differ by $< 7.5$\,mm have their unaries blended with weight $0.45 \exp(-\Delta z^2 / 32)$.

For $N$ candidates sorted by decreasing $z_{\mathrm{sort}}$, the Viterbi table $D(i, k)$ (minimum cost of assigning candidate $i \in \{1,\dots,N\}$ to label $k \in \{1,\dots,12\}$ over strictly increasing indices with at most two skipped labels, $\Delta k = k - k' \in \{1, 2, 3\}$) satisfies
\begin{equation}
  \begin{aligned}
    D(i, k) &= U(i, k) + \min\!\Bigg( 3(k-1) + \sum_{m=1}^{i-1}\kappa_m, \\
    &\qquad \min_{\substack{1 \le i' < i \\ \max(1, k-3) \le k' < k}} \!\bigg[ D(i', k') + \sum_{m=i'+1}^{i-1}\kappa_m + \Psi(i', k' \!\to\! i, k) \bigg]\!\Bigg),
  \end{aligned}
  \label{eq:viterbi}
\end{equation}
and the optimal assignment minimizes $D(i, k) + 1.5(12 - k) + \sum_{m=i+1}^{N}\kappa_m$ (or the all-clutter cost $\sum_{m=1}^{N}\kappa_m + 36$). Skipping candidate $m$ as clutter costs $\kappa_m = 65$ if $L_m \ge 115$\,mm, $22$ if $L_m \ge 55$\,mm and $\theta_{\min,m} < -0.45$\,rad, and $2.5 + 8\,\mathrm{clip}((L_m - 25)/50, 0, 1)$ otherwise. With $\Delta z = z_{\mathrm{sort},i'} - z_{\mathrm{sort},i}$, cumulative spacing $\mu_{k'\to k} = \sum_{m=k'}^{k-1}\mu_{s,m}$, and cumulative transition variance $\sigma_{k'\to k}^2 = s_Z^2 \sum_{m=k'}^{k-1}\sigma_{s,m}^2 + 6$, transitions require $\Delta z \ge 4.5$\,mm and cost $\Psi(i', k' \!\to\! i, k) = (\Delta z - s_Z \mu_{k' \to k})^2 / (2\,\sigma_{k' \to k}^2) + \gamma_{\mathrm{skip}}(\Delta k - 1)$, where $\gamma_{\mathrm{skip}} = 2.5$ if $\Delta z \ge 1.55\, s_Z \mu_{s,k'}$ and $18$ otherwise. After a first solution per side, a second pass subtracts $3 \exp(-\Delta z^2/32)$ from $U(i,k)$ whenever the contralateral rib labeled $k$ lies within $8.5$\,mm, and Eq.~\ref{eq:viterbi} is re-solved.

\subsection{\texttt{RibSnake} Refinement}
(1) \emph{Apical trimming}: an \texttt{R01}/\texttt{L01} longer than $135$\,mm is cut where $|X - x_{\mathrm{spine}}|$ first exceeds $0.55\,W$ after index 6 (if the remainder is $\ge 45$\,mm), and if the resulting curve still exceeds $140$\,mm it is truncated at $135$\,mm arc length.
(2) \emph{Terminal-rib recovery}: a missing twelfth rib is searched in a box $22$--$44$\,mm below the posterior end of rib 11 for voxels with $P \ge 0.30$ and $\max_{3^3} I \ge 150$\,HU at least $18$\,mm from rib 11, accepted if $\ge 12$ voxels with $\max P \ge 0.45$ exist and the contralateral twelfth rib is present or also recovered; a missing first or twelfth rib is otherwise synthesized from the patient-conditioned template (atlas curve displaced by a smoothed interpolation of the displacements observed on the detected ribs, with mirrored contralateral coupling), anchored in $Z$ to the contralateral rib, kept only if $\ge 11$\,mm from the adjacent rib, and snapped point-wise to the probability-weighted bone centroid in a $7^3$-voxel box when the local score $P\,\mathrm{clip}((G(I) - 60)/300, 0, 1.5)$ reaches $0.14$.
(3) \emph{Normal-plane centering}: every curve is resampled to $N = \max(15, \lfloor L / 2\,\text{mm} \rceil)$ points; for 5 iterations each point $\mathbf{p}_m$ moves by $0.35\times$ the offset of the weighted centroid on a $5 \times 5$ local normal-plane grid $\mathbf{x}_{u,v} = \mathbf{p}_m + u\mathbf{n}_1 + v\mathbf{n}_2$ (step $1.25$\,mm, $u^2+v^2 \le 6.5\,\text{mm}^2$) with weights $w(u,v) = \mathrm{clip}((G_{1.0}(I)(\mathbf{x}_{u,v}) - 40)/350, 0, 2)\,(0.25 + 0.75\,\mathrm{clip}(P_{\mathrm{hyb}}(\mathbf{x}_{u,v}), 0, 1))$, is smoothed along the curve ($\sigma = 1.2$ points, weight $0.45$), and is clamped to $1.6$\,mm total displacement; finally a cubic smoothing spline ($s = 0.15\,N$) is fitted and re-evaluated at $N$ points. Output curves therefore have between 15 and roughly 200 points (median $\approx 150$), not a fixed count.

\section{Experimental Results}
\label{sec:experiments}

\subsection{Data, Protocol and Metrics}
We use the SCR 3D thoracic CT rib centerline dataset \cite{wu2012learning}: 60 anonymized scans resampled to $2$\,mm isotropic voxels (typical size $155 \times 142 \times 165$) with manually annotated, B-spline-interpolated centerlines. The split (\texttt{split.json}) is the one used throughout development: 48 training scans (47 with annotations; $1{,}116$ ribs) and 12 evaluation scans. It is \emph{not} random: the 12 evaluation scans are the case-sensitive alphabetical tail of the file list. One evaluation file (\texttt{RIB\_007}) labels two ribs as \texttt{R07} (640 points, a 220\,mm jump between two curves at the heights of \texttt{L07} and \texttt{L08}); the released script \texttt{fix\_gt\_annotations} splits it into \texttt{R07} and \texttt{R08} (the only file changed, logged in \texttt{fix\_log.json}), giving $286$ evaluation ribs (10 scans with 24, \texttt{Tsc\_004002} with 22). Results on the unmodified ground truth are also reported.

\textbf{Protocol caveat.} The $\approx 60$ numerical constants of Stage~3 (Sec.~\ref{sec:stage3}) were tuned by repeatedly running the pipeline and inspecting metrics on these 12 scans; the three learned detectors and the atlas, in contrast, were fitted on the 47 training scans only. The 12-scan results below are therefore optimistic for Stage~3 and should be read as in-sample for its constants. Section~\ref{sec:cv} quantifies the effect with a 5-fold cross-validation in which the detectors and the atlas are re-fitted per fold and the Stage-3 constants are frozen.

\textbf{Metrics} (implemented in \texttt{ribline.model.metrics}). For a ground-truth rib $r$ and the prediction $\hat r$ with the same label let $d_{\mathrm{p2g}}$ and $d_{\mathrm{g2p}}$ be the mean prediction-to-GT and GT-to-prediction point-to-curve distances.
(1) \textbf{Labeled symmetric Chamfer distance} (LCD, mm): $\tfrac{1}{2}(d_{\mathrm{p2g}} + d_{\mathrm{g2p}})$ averaged over \emph{all} GT ribs, a missing label counting $50$\,mm.
(2) \textbf{Unlabeled Chamfer distance} (UCD) and HD95 on the pooled point clouds.
(3) \textbf{Labeled clDice} \cite{shit2021cldice} at $3$ and $4$\,mm, averaged over GT ribs (missing $= 0$); unlabeled clDice is also computed.
(4) \textbf{Labeling accuracy}: a GT rib is correct if $\hat r$ exists, $d_{\mathrm{p2g}} \le 10$\,mm \emph{and} $d_{\mathrm{g2p}} \le 10$\,mm (symmetric criterion), and $r$ is the GT rib closest to $\hat r$ (up to a $1.5$\,mm tie-breaking tolerance on $d_{\mathrm{p2g}}$); reported pooled over ribs for all ribs ($\mathrm{Acc}_A$), first ($\mathrm{Acc}_F$, $n = 24$), intermediate ($\mathrm{Acc}_I$, $n = 240$) and twelfth ribs ($\mathrm{Acc}_T$, $n = 22$).
(5) \textbf{Recall}: a GT rib is detected if any predicted curve satisfies the symmetric criterion. (6) \textbf{Precision}: fraction of predicted curves with $d_{\mathrm{p2g}} \le 10$\,mm to some GT rib (its complement is the false-positive curve count; we also compute symmetric curve precision requiring both $d_{\mathrm{p2g}} \le 10$\,mm and $d_{\mathrm{g2p}} \le 10$\,mm). (7) \textbf{Normalized length difference} (NLD) $|L(\hat r) - L(r)| / L(r)$.
Subject-level means are given with standard deviations and 95\,\% bootstrap confidence intervals over subjects ($10{,}000$ resamples); pooled rates with Wilson 95\,\% intervals; differences between methods with two-sided Wilcoxon signed-rank tests over the 12 subjects.

\subsection{Ablation on the 12 Evaluation Scans}
All rows of Table~\ref{tab:ablation} are regenerated from the stored probability maps with the released Stage-3 code (\texttt{ribline.scripts.run\_ablation}); ``Z-sort'' denotes the same extraction and consolidation followed by labeling the candidates of each side $1..12$ in decreasing $z_{\mathrm{sort}}$ instead of the Viterbi program, without \texttt{RibSnake}.

\begin{table*}[t]
\centering
\caption{Ablation on the 12 evaluation scans ($286$ GT ribs, RIB\_007 annotation fixed). LCD: labeled symmetric Chamfer distance with 50\,mm penalty for missed labels (subject mean $\pm$ SD and 95\,\% bootstrap CI); clDice: labeled, 3\,mm / 4\,mm; accuracies, recall and precision pooled over ribs (symmetric 10\,mm criterion); FP: false-positive curves out of all predicted curves. Note the protocol caveat: Stage-3 constants were developed on these scans.}
\label{tab:ablation}
\resizebox{\textwidth}{!}{%
\setlength{\tabcolsep}{3.5pt}
\begin{tabular}{l|cc|c|cc|cccc|cc|c}
\toprule
\multirow{2}{*}{\textbf{Method}} & \textbf{LCD (mm)} $\downarrow$ & \textbf{95\,\% CI} & \textbf{UCD} $\downarrow$ & \multicolumn{2}{c|}{\textbf{lab.\ clDice} $\uparrow$} & \multicolumn{4}{c|}{\textbf{Labeling accuracy (\%)} $\uparrow$} & \textbf{Recall} $\uparrow$ & \textbf{Prec.\ (FP)} $\uparrow$ & \textbf{NLD} \\
 & mean $\pm$ SD & & (mm) & 3\,mm & 4\,mm & $\mathrm{Acc}_A$ & $\mathrm{Acc}_F$ & $\mathrm{Acc}_I$ & $\mathrm{Acc}_T$ & (\%) & (\%) & $\downarrow$ \\
\midrule
RF Pass-1 + Z-sort & $14.10 \pm 11.10$ & $[7.96, 20.46]$ & $3.31$ & $0.548$ & $0.606$ & $64.3$ & $79.2$ & $62.9$ & $63.6$ & $91.3$ & $95.5$ (13) & $0.264$ \\
SCF Pass-2 + Z-sort & $21.30 \pm 13.33$ & $[13.79, 28.67]$ & $4.16$ & $0.416$ & $0.452$ & $43.0$ & $45.8$ & $44.2$ & $27.3$ & $76.2$ & $100.0$ (0) & $0.254$ \\
SCF Pass-2 + Stage 3 & $9.36 \pm 10.57$ & $[4.24, 16.19]$ & $3.73$ & $0.683$ & $0.720$ & $71.0$ & $54.2$ & $73.8$ & $59.1$ & $78.0$ & $99.6$ (1) & $0.226$ \\
Anatomical SCF + Stage 3 & $2.88 \pm 1.32$ & $[2.23, 3.73]$ & $2.41$ & $0.845$ & $0.915$ & $95.1$ & $91.7$ & $95.0$ & $100.0$ & $95.1$ & $99.7$ (1) & $0.088$ \\
3D U-Net + Z-sort & $7.57 \pm 7.43$ & $[3.97, 12.17]$ & $2.75$ & $0.710$ & $0.813$ & $83.6$ & $70.8$ & $86.7$ & $63.6$ & $92.0$ & $100.0$ (0) & $0.147$ \\
3D U-Net + Stage 3 & $3.40 \pm 1.99$ & $[2.42, 4.64]$ & $2.30$ & $0.835$ & $0.902$ & $92.0$ & $66.7$ & $95.4$ & $81.8$ & $93.0$ & $100.0$ (0) & $0.121$ \\
Hybrid + Viterbi (w/o RibSnake) & $2.85 \pm 1.70$ & $[2.07, 3.95]$ & $2.11$ & $0.872$ & $0.933$ & $95.8$ & $83.3$ & $97.5$ & $90.9$ & $95.8$ & $99.6$ (1) & $0.147$ \\
\textbf{Hybrid + full Stage 3 (ours)} & $\mathbf{2.05 \pm 0.48}$ & $[1.80, 2.33]$ & $\mathbf{1.91}$ & $\mathbf{0.886}$ & $\mathbf{0.945}$ & $\mathbf{97.2}$ & $\mathbf{91.7}$ & $\mathbf{97.5}$ & $\mathbf{100.0}$ & $\mathbf{97.2}$ & $\mathbf{100.0}$ (0) & $\mathbf{0.068}$ \\
\bottomrule
\end{tabular}%
}
\end{table*}

\begin{figure}[t]
  \centering
  \includegraphics[width=\textwidth]{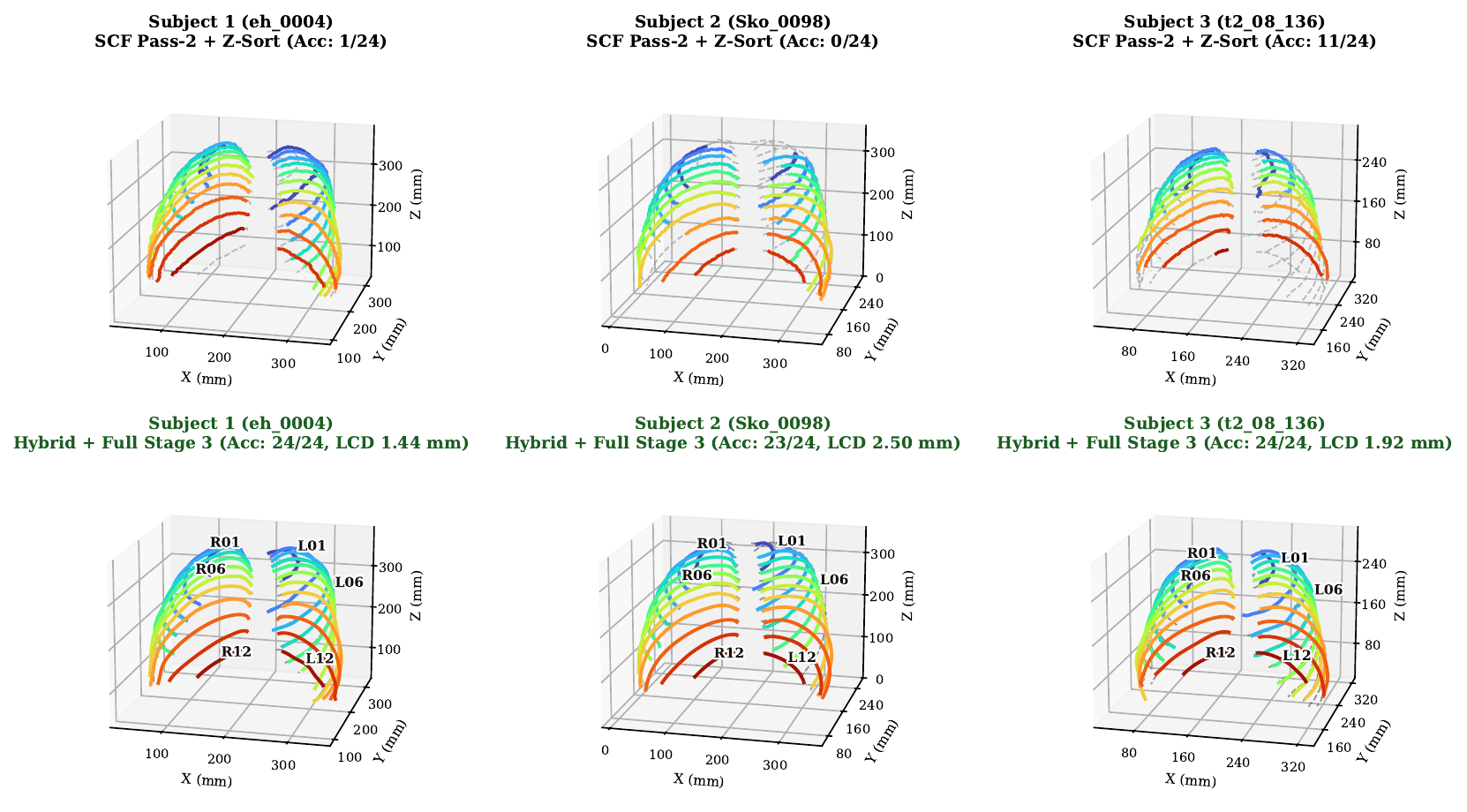}
  \vspace{-2mm}
  \caption{Three evaluation scans (dashed gray: ground truth; colored: predicted labels 1--12). Top: SCF Pass-2 + Z-sort baseline. Bottom: hybrid + full Stage~3; titles give the number of correctly labeled ribs under the symmetric 10\,mm criterion and the LCD.}
  \label{fig:3d_centerlines}
  \vspace{-2mm}
\end{figure}

\begin{figure}[t]
  \centering
  \includegraphics[width=\textwidth]{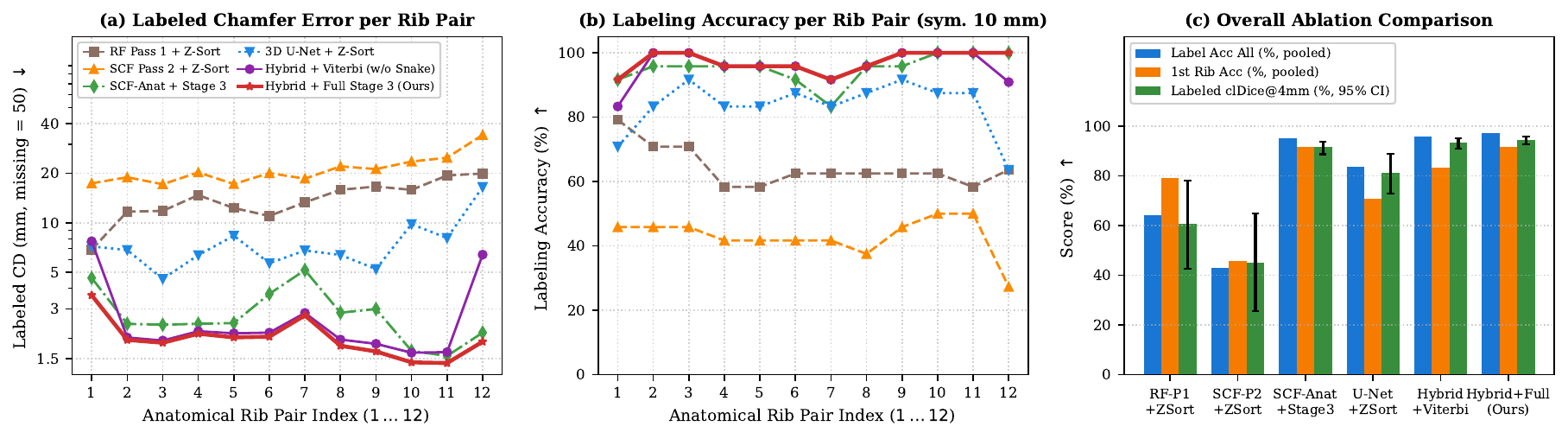}
  \vspace{-2mm}
  \caption{Per-rib analysis on the 12 evaluation scans: (a) LCD per rib pair (log scale, missing labels $=50$\,mm); (b) labeling accuracy per rib pair; (c) pooled $\mathrm{Acc}_A$, $\mathrm{Acc}_F$ and labeled clDice@4\,mm with 95\,\% bootstrap CIs.}
  \label{fig:metrics_plot}
  \vspace{-2mm}
\end{figure}

\textbf{Observations.} Replacing craniocaudal counting by the Stage-3 labeler is the largest single effect: for the U-Net it reduces LCD from $7.57$ to $3.40$\,mm and raises $\mathrm{Acc}_A$ from $83.6$ to $92.0\,\%$ (Wilcoxon on LCD $p = 0.001$); for the Pass-2 forest from $21.30$ to $9.36$\,mm. Notably, Pass-2 auto-context recall drops out-of-sample relative to Pass-1 ($91.3 \to 76.2\,\%$ with Z-sort, $78.0\,\%$ with Stage~3) because its long-range ($\pm 80$\,mm) probability-difference rays overfit to the sharp in-sample Pass-1 training probability maps without explicit canonical coordinates $(u,v,w)$ (compounded by a slight Pass-1 training-pass provenance shift between \texttt{full\_v4} and \texttt{full\_v5}). Conversely, the anatomically anchored SCF alone already reaches $2.88$\,mm and $95.1\,\%$, and is clearly better than both the Pass-2 forest and the standalone 3D U-Net on first and twelfth ribs: because the 3D U-Net operates on translation-invariant $64^3$ local patches lacking global canonical coordinates $(u,v,w)$, its accuracy drops on apical first and floating twelfth ribs ($66.7\,\%$ and $81.8\,\%$ on the 12 scans; $61.5\,\%$ and $67.8\,\%$ in 5-fold CV), which max-fusing with the coordinate-aware Anatomical SCF resolves ($91.7\,\%$ and $100.0\,\%$ on the 12 scans; $82.9\,\%$ and $96.5\,\%$ in 5-fold CV). Fusing SCF and U-Net improves every metric over either detector, and \texttt{RibSnake} further reduces LCD from $2.85$ to $2.05$\,mm ($p = 0.009$, subject-level median LCD $1.89$\,mm) and NLD from $0.147$ to $0.068$ ($p = 0.0005$), mainly by trimming and completing the terminal ribs ($\mathrm{Acc}_F$ $83.3 \to 91.7\,\%$, $\mathrm{Acc}_T$ $90.9 \to 100\,\%$); its effect on labeled clDice ($0.872 \to 0.886$) and on $\mathrm{Acc}_A$ ($274 \to 278$ ribs) is not significant on 12 subjects ($p = 0.20$ and $p = 0.50$).

The final system produces exactly 24 curves on every scan with 24 annotated ribs, 22 on \texttt{Tsc\_004002}, and no false-positive curve ($286/286$ predicted curves lie on a GT rib). The eight ribs that fail the symmetric criterion are all \emph{partially traced} ribs rather than mislabeled ones ($d_{\mathrm{p2g}} \le 4.2$\,mm but $d_{\mathrm{g2p}}$ of $10.6$--$34.8$\,mm): \texttt{L07} of \texttt{Sko\_004029} (93 predicted vs.\ 291 GT points, HD95 $66$\,mm), \texttt{L04}--\texttt{L08} of \texttt{Zer\_004061} (anterior third missing after a costal-cartilage gap) and both first ribs of \texttt{pat1}. Under the legacy one-sided criterion all 286 ribs would count as correct; we consider the symmetric figure the honest one. Per rib, 25 of 286 have HD95 $> 10$\,mm (8 above $20$\,mm) and 52 have clDice@3\,mm $< 0.8$ (5 below $0.5$). Per-subject LCD ranges from $1.44$ (\texttt{pet\_ct\_bc\_\_eh}) to $2.98$\,mm (\texttt{Zer\_004061}); labeled HD95 averages $5.1$\,mm.

\textbf{Sensitivity to the ground truth and to a possible duplicate.} On the unmodified annotations ($285$ ribs, merged \texttt{R07} kept) the final system scores $2.08 \pm 0.46$\,mm ($[1.83, 2.35]$) and $96.8\,\%$ ($276/285$); excluding the merged rib instead ($284$ ribs) gives $2.05 \pm 0.48$\,mm and $97.2\,\%$ ($276/284$). The training scan \texttt{Bro\_004029\_0040290098} and the evaluation scan \texttt{Sko\_004029\_0040290098} share an accession-like identifier; they have different matrix sizes, intensity encodings and fields of view (voxel correlation $0.25$ on the common crop), but their per-rib arc-length fingerprints are among the 1\,\% most similar of all $1{,}653$ scan pairs, so we cannot exclude that they are two series of one patient. Dropping \texttt{Sko} changes the final result to $2.01 \pm 0.48$\,mm ($[1.75, 2.31]$) and $97.3\,\%$ ($255/262$).

\subsection{Cross-Validation with Frozen Stage-3 Constants}
\label{sec:cv}
To quantify how much the development protocol favored the 12 evaluation scans we run two experiments.

\textbf{(a) Frozen pipeline on the 47 training scans.} The released detectors, atlas and Stage-3 code are applied unchanged to the 47 annotated training scans ($1{,}116$ ribs). The detectors are in-sample here, but the Stage-3 constants were never inspected on these scans, so a drop relative to Table~\ref{tab:ablation} isolates the Stage-3 over-tuning plus the larger anatomical variety of the training cohort. Table~\ref{tab:train47} gives the result: the full pipeline reaches an LCD of $2.63 \pm 1.91$\,mm (CI $[2.14, 3.23]$; median $1.89$\,mm), labeled clDice $0.885 / 0.932$, $\mathrm{Acc}_A = 95.4\,\%$ ($1{,}065/1{,}116$; Wilson $94.0$--$96.5\,\%$), recall $95.6\,\%$ and precision $98.2\,\%$ (20 false-positive curves out of $1{,}126$), with 2 missed labels; 28 of the 47 scans are labeled perfectly and 4 have an LCD above $5$\,mm (worst: \texttt{HP\_485533} $10.3$\,mm, \texttt{03\_Anonymous\_0002} $9.9$\,mm, both with label shifts on one hemithorax). The ranking of the ablation rows is preserved (hybrid $>$ SCF $>$ U-Net; \texttt{RibSnake} reduces LCD $2.98 \to 2.63$\,mm and NLD $0.155 \to 0.083$, Wilcoxon $p < 10^{-5}$ on 47 subjects, without changing $\mathrm{Acc}_A$, $p = 0.61$), but the absolute level is clearly below the 12-scan figures ($2.05$\,mm / $97.2\,\%$). We therefore regard the 12-scan numbers as an optimistic upper bound and the training-scan numbers as a more realistic indication of what the released constants deliver on SCR-like data; the mean LCD is dominated by a few scans with label shifts (the median is $1.89$\,mm, identical to the 12-scan evaluation median).

\begin{table}[t]
\centering
\caption{Frozen pipeline (released detectors, atlas, and Stage-3 constants) on the 47 annotated training scans ($1{,}116$ GT ribs; detectors in-sample, Stage-3 constants never tuned on these scans). Same metric definitions as Table~\ref{tab:ablation}.}
\label{tab:train47}
\resizebox{\textwidth}{!}{%
\setlength{\tabcolsep}{3.2pt}
\begin{tabular}{l|cc|c|cc|cccc|cc|c}
\toprule
\textbf{Method} & \textbf{LCD (mm)} $\downarrow$ & \textbf{95\,\% CI} & \textbf{UCD (mm)} $\downarrow$ & \multicolumn{2}{c|}{\textbf{lab.\ clDice} $\uparrow$} & $\mathrm{Acc}_A\uparrow$ & $\mathrm{Acc}_F\uparrow$ & $\mathrm{Acc}_I\uparrow$ & $\mathrm{Acc}_T\uparrow$ & \textbf{Recall} $\uparrow$ & \textbf{Prec.\ (FP)} $\uparrow$ & \textbf{NLD} $\downarrow$ \\
\midrule
Anatomical SCF + Stage 3 & $3.62 \pm 3.97$ & $[2.67, 4.91]$ & $2.97$ & $0.830$ & $0.895$ & $92.4$ & $89.2$ & $92.4$ & $95.7$ & $94.8$ & $97.4$ (29) & $0.094$ \\
3D U-Net + Stage 3 & $5.84 \pm 7.20$ & $[3.95, 8.03]$ & $2.98$ & $0.785$ & $0.826$ & $83.9$ & $69.9$ & $86.0$ & $76.3$ & $91.6$ & $97.8$ (25) & $0.142$ \\
Hybrid + Viterbi (w/o RibSnake) & $2.98 \pm 2.11$ & $[2.42, 3.63]$ & $2.77$ & $0.875$ & $0.930$ & $\mathbf{95.6}$ & $\mathbf{88.2}$ & $\mathbf{96.1}$ & $\mathbf{97.8}$ & $\mathbf{96.1}$ & $\mathbf{98.5}$ (17) & $0.155$ \\
\textbf{Hybrid + full Stage 3 (ours)} & $\mathbf{2.63 \pm 1.91}$ & $[2.14, 3.23]$ & $\mathbf{2.62}$ & $\mathbf{0.885}$ & $\mathbf{0.932}$ & $95.4$ & $86.0$ & $\mathbf{96.1}$ & $\mathbf{97.8}$ & $95.6$ & $98.2$ (20) & $\mathbf{0.083}$ \\
\bottomrule
\end{tabular}%
}
\end{table}

\textbf{(b) 5-fold cross-validation with frozen Stage-3 constants.} The 59 annotated scans ($1{,}402$ GT ribs) are dealt into five seeded random folds of sizes 12/12/12/12/11 (\texttt{cv5/folds.json}); for each fold the 24-rib statistical atlas, the anatomically anchored SCF (150 trees, depth 24, $50{,}000$ samples/scan) and the 3D U-Net (20 epochs, 8 patches/scan/epoch) are re-fitted from scratch on the fold's training scans, and the released Stage-3 code with frozen constants is evaluated on the held-out fold (\texttt{ribline.scripts.run\_cv5} / \texttt{experiments/rebuttal/cv\_fold.py}, pooled in \texttt{benchmark\_cv5.json}). Table~\ref{tab:cv5} reports the pooled 59-scan out-of-fold results: the full hybrid pipeline achieves an LCD of $3.64 \pm 3.90$\,mm (95\,\% bootstrap CI $[2.79, 4.74]$, median $2.39$\,mm; per-fold mean LCDs $3.12, 4.80, 4.91, 2.84, 2.43$\,mm), UCD $2.85$\,mm, labeled clDice $0.830$ (3\,mm) / $0.889$ (4\,mm), symmetric labeling accuracy $\mathrm{Acc}_A = 91.8\,\%$ ($1{,}287/1{,}402$, Wilson 95\,\% CI $90.2$--$93.1\,\%$; $\mathrm{Acc}_F = 82.9\,\%$, $\mathrm{Acc}_I = 92.2\,\%$, $\mathrm{Acc}_T = 96.5\,\%$), recall $94.6\,\%$, precision $97.7\,\%$ (33 false-positive curves out of $1{,}412$ predicted) and NLD $0.097$. Thirty of the 59 scans achieve $100\,\%$ symmetric labeling accuracy, while 8 scans have LCD $> 5$\,mm (worst: \texttt{BC3\_Anonymous\_0001} at $26.58$\,mm and \texttt{BC1\_Anonymous\_0001} at $15.76$\,mm, where superior CT field-of-view truncation clamps $Z_{\mathrm{apex}}$ to the top slice or admits an unmerged apical clavicle fragment, shifting relative heights $z_{\mathrm{post}} = Z_{\mathrm{post}} - Z_{\mathrm{apex}}$ by one intercostal space, $\sim 22$\,mm). Across all $n = 59$ held-out subjects, hybrid fusion significantly outperforms both the standalone Anatomical SCF (Wilcoxon $p = 2.1 \times 10^{-7}$ on LCD, $p = 9.0 \times 10^{-5}$ on $\mathrm{Acc}_A$) and the standalone 3D U-Net ($p = 1.9 \times 10^{-5}$ on LCD, $p = 8.5 \times 10^{-6}$ on $\mathrm{Acc}_A$), and \texttt{RibSnake} significantly improves LCD ($4.17 \to 3.64$\,mm, $p = 2.0 \times 10^{-9}$), labeled clDice@3\,mm ($0.807 \to 0.830$, $p = 3.2 \times 10^{-9}$) and NLD ($0.195 \to 0.097$, $p = 4.4 \times 10^{-11}$). Comparing the three evaluation regimes---$2.05$\,mm / $97.2\,\%$ on the 12-scan split (detectors out-of-sample, Stage~3 in-sample), $2.63$\,mm / $95.4\,\%$ on the 47 training scans (detectors in-sample, Stage~3 out-of-sample), and $3.64$\,mm / $91.8\,\%$ in 5-fold CV (both detectors and Stage~3 out-of-sample, median LCD $2.39$\,mm)---cleanly decomposes the optimism of the original 12-scan benchmark while confirming that every architectural contribution (anatomical SCF, hybrid max-fusion, bilateral Viterbi, and \texttt{RibSnake}) holds out-of-sample with high statistical significance.

\begin{table}[t]
\centering
\caption{5-fold cross-validation across all 59 annotated SCR scans ($1{,}402$ GT ribs; per-fold re-fitted atlas, Anatomical SCF, and 3D U-Net; frozen Stage-3 constants; \texttt{benchmark\_cv5.json}).}
\label{tab:cv5}
\resizebox{\textwidth}{!}{%
\setlength{\tabcolsep}{3.2pt}
\begin{tabular}{l|cc|c|cc|cccc|cc|c}
\toprule
\textbf{Method} & \textbf{LCD (mm)} $\downarrow$ & \textbf{95\,\% CI} & \textbf{UCD (mm)} $\downarrow$ & \multicolumn{2}{c|}{\textbf{lab.\ clDice} $\uparrow$} & $\mathrm{Acc}_A\uparrow$ & $\mathrm{Acc}_F\uparrow$ & $\mathrm{Acc}_I\uparrow$ & $\mathrm{Acc}_T\uparrow$ & \textbf{Recall} $\uparrow$ & \textbf{Prec.\ (FP)} $\uparrow$ & \textbf{NLD} $\downarrow$ \\
\midrule
Anatomical SCF + Stage 3 & $5.28 \pm 4.62$ & $[4.16, 6.56]$ & $3.44$ & $0.769$ & $0.840$ & $86.9$ & $72.6$ & $87.7$ & $93.0$ & $90.8$ & $95.3$ (67) & $0.126$ \\
3D U-Net + Stage 3 & $7.44 \pm 8.93$ & $[5.31, 9.85]$ & $3.30$ & $0.745$ & $0.785$ & $78.2$ & $61.5$ & $80.9$ & $67.8$ & $87.0$ & $\mathbf{98.1}$ (26) & $0.170$ \\
Hybrid + Viterbi (w/o RibSnake) & $4.17 \pm 4.01$ & $[3.29, 5.28]$ & $3.07$ & $0.807$ & $0.879$ & $91.4$ & $81.2$ & $\mathbf{92.2}$ & $93.9$ & $94.3$ & $97.9$ (29) & $0.195$ \\
\textbf{Hybrid + full Stage 3 (ours)} & $\mathbf{3.64 \pm 3.90}$ & $[2.79, 4.74]$ & $\mathbf{2.85}$ & $\mathbf{0.830}$ & $\mathbf{0.889}$ & $\mathbf{91.8}$ & $\mathbf{82.9}$ & $\mathbf{92.2}$ & $\mathbf{96.5}$ & $\mathbf{94.6}$ & $97.7$ (33) & $\mathbf{0.097}$ \\
\bottomrule
\end{tabular}%
}
\end{table}

\subsection{Relation to Published Numbers}
Wu \textit{et al.} \cite{wu2012learning} report $89.5$--$95.9\,\%$ labeling accuracy on an SCR cohort, Lenga \textit{et al.} \cite{lenga2018} $94.1\,\%$ on their own data, and RibSeg~v2 \cite{ribseg2024} an LCD of $3.15$\,mm and $96.5\,\%$ label accuracy on sub-millimeter RibFrac CT. These numbers come from different scans, resolutions, splits and (in particular) matching criteria, and we did not re-run those methods; we therefore make \emph{no} claim of outperforming them. Our contribution is a documented, reproducible open-source pipeline whose accuracy on this 2\,mm benchmark is in the same range as the published systems, together with an evaluation protocol that we hope will make such comparisons possible in the future.

\section{Limitations}
\label{sec:limitations}
\begin{itemize}
  \item \textbf{Dataset size and provenance.} The SCR benchmark contains 60 scans from a single source, resampled to $2$\,mm; the field of view, kernel and intensity encoding vary between scans but no acquisition metadata are available, so no sub-group analysis is possible and generalization to other scanners, resolutions or pediatric/pathological anatomy is untested.
  \item \textbf{Evaluation set.} Only 12 scans ($286$ ribs) are held out, and the split is the alphabetical tail of the file list rather than a random draw; with 8 errors in 286 ribs the Wilson interval for $\mathrm{Acc}_A$ is $94.6$--$98.6\,\%$, and 0 of 12 subject-level failures only bounds the per-scan failure rate below $\approx 25\,\%$ (rule of three).
  \item \textbf{Stage-3 constants were tuned with access to the evaluation scans.} We disclose this and report the cross-validation of Sec.~\ref{sec:cv}; a truly untouched test set does not exist for this dataset, and the released constants should be regarded as fitted to SCR.
  \item \textbf{Annotation quality.} The ground truth contains at least one merged-rib error (\texttt{RIB\_007}) and the twelfth ribs are annotated inconsistently (absent in \texttt{Tsc\_004002}, very short in others); a fraction of the residual error is annotation noise that we cannot separate from algorithmic error.
  \item \textbf{Possible patient overlap} between \texttt{Bro\_004029} (training) and \texttt{Sko\_004029} (evaluation) cannot be excluded from the anonymized data.
  \item \textbf{Partial traces and superior FOV truncation.} On the 12 evaluation scans, the dominant failure mode is a rib traced only over its posterior part across a costal-cartilage gap; the symmetric criterion counts these as errors while the one-sided criterion does not. In 5-fold cross-validation across all 59 scans, the 8 outlier scans with LCD $> 5$\,mm (e.g.\ \texttt{BC3\_Anonymous\_0001} and \texttt{BC1\_Anonymous\_0001}) are primarily caused by superior CT field-of-view truncation, which clamps $Z_{\mathrm{apex}}$ to the top axial slice or admits an unmerged apical clavicle fragment and shifts relative heights $z_{\mathrm{post}} = Z_{\mathrm{post}} - Z_{\mathrm{apex}}$ by one intercostal space ($\sim 22$\,mm). Precision is reported at the curve level only.
\end{itemize}

\section{Conclusion}
\label{sec:conclusion}
We presented a three-stage 3D rib centerline extraction and anatomical labeling pipeline that combines a lung-apex and vertebral-midline thoracic coordinate frame, a hybrid Semantic Context Forest + 3D U-Net detector, and a patient-scaled bilateral Viterbi labeler with deformable refinement (\texttt{RibSnake}). On the SCR 2\,mm thoracic CT benchmark, the framework achieves $2.05 \pm 0.48$\,mm labeled Chamfer distance and $97.2\,\%$ symmetric labeling accuracy on the 12 evaluation scans, $2.63 \pm 1.91$\,mm ($95.4\,\%$, median $1.89$\,mm) with frozen Stage-3 constants on the 47 training scans, and $3.64 \pm 3.90$\,mm ($91.8\,\%$, median $2.39$\,mm) in 5-fold cross-validation across all 59 annotated scans. Our ablation and cross-validation analyses demonstrate that coupling an anatomically anchored forest with a lightweight 3D U-Net improves terminal-rib recall and shaft continuity over either detector alone, while bilateral polar dynamic programming and normal-plane active contour refinement substantially reduce index shifts and terminal-rib truncation. Details of the open-source Python library and pretrained model checkpoints are provided in Appendix~\ref{app:github} and Appendix~\ref{app:huggingface}.

\clearpage
\appendix

\section{Open-Source \texttt{ribline} Library on GitHub}
\label{app:github}

To support transparent evaluation and reproducible research in 3D rib centerline extraction and anatomical numbering, we release the complete \texttt{ribline} Python library (\texttt{v0.3.1}) under the MIT License at \url{https://github.com/wq2012/ribline}.

\subsection{Package Architecture and Module Mapping}
The repository is structured as a self-contained, pip-installable Python package (\texttt{pyproject.toml}) backed by unit tests (\texttt{run\_tests.py}) and GitHub Actions continuous integration. Table~\ref{tab:app_modules} maps each module in the open-source library to the corresponding methodological sections and equations in this paper.

\begin{table}[h!]
\vspace{-2mm}
\centering
\caption{Mapping between modules in the open-source \texttt{ribline} library (\url{https://github.com/wq2012/ribline}) and the sections of this paper.}
\label{tab:app_modules}
\resizebox{\textwidth}{!}{%
\footnotesize
\setlength{\tabcolsep}{4pt}
\begin{tabular}{llp{6.6cm}}
\toprule
\textbf{Module / Script} & \textbf{Section} & \textbf{Implemented Functionality} \\
\midrule
\texttt{ribline.data.image\_reader} & Sec.~\ref{sec:frame} & Analyze 7.5 (\texttt{.hdr/.img}), NIfTI (\texttt{.nii/.nii.gz}), and MetaImage (\texttt{.mha/.mhd}) 3D I/O with spacing and origin preservation. \\
\texttt{ribline.data.centerline\_parser} & Sec.~\ref{sec:experiments} & Parsing of ground-truth (\texttt{\_interp.xls}) and predicted 3D centerline coordinate files. \\
\texttt{ribline.model.anatomy} & Sec.~\ref{sec:frame} & Hounsfield standardization (\texttt{standardize\_hu}), lung apex and spine midline detection, kyphosis-aware azimuthal angle, and 24-rib statistical atlas (\texttt{RibCageStatisticalAtlas}). \\
\texttt{ribline.model.features} & Sec.~\ref{sec:detection} & Multi-scale 3D Gaussian derivatives, diagonal-curvature tubularity proxy, canonical coordinates, and Random Shift Intensity/Probability Differences (RSID). \\
\texttt{ribline.model.forest} & Sec.~\ref{sec:detection} & Balanced random forest wrappers (\texttt{RibForest}) for Pass-1, Pass-2 auto-context, and anatomically anchored SCF detectors. \\
\texttt{ribline.model.unet.model} & Sec.~\ref{sec:detection} & Plain 3-level 3D U-Net (\texttt{RibUNet3D}, $339{,}889$ parameters) with GroupNorm and sliding-window inference. \\
\texttt{ribline.model.reconstruction} & Sec.~\ref{sec:stage3} & Anatomical corridor masking, 3D medial-axis skeletonization, junction splitting, and azimuthal fragment chaining (\texttt{CenterlineExtractor}). \\
\texttt{ribline.model.topology} & Sec.~\ref{sec:stage3} & Candidate consolidation, patient scale estimation $s_Z$, and contralateral-coupled bilateral Viterbi labeling (\texttt{BilateralViterbiLabeler}). \\
\texttt{ribline.model.refinement} & Sec.~\ref{sec:stage3} & Apical clavicle trimming, patient-conditioned template recovery for terminal ribs, and normal-plane active contour centering (\texttt{RibSnake}). \\
\texttt{ribline.model.metrics} & Sec.~\ref{sec:experiments} & Symmetric 10\,mm labeling accuracy, 50\,mm missed-rib Labeled Chamfer Distance (LCD), unlabeled CD/HD95, labeled clDice at 3/4\,mm, recall, precision, and NLD. \\
\texttt{ribline.hub} & App.~\ref{app:huggingface} & Programmatic Hugging Face Hub checkpoint downloading and model instantiation (\texttt{download\_pretrained\_models}, \texttt{load\_pretrained\_models}). \\
\bottomrule
\end{tabular}%
}
\vspace{-3mm}
\end{table}

\subsection{End-to-End Command-Line Workflow}
The \texttt{ribline.scripts} subpackage provides command-line entry points to reproduce every stage of the pipeline from raw thoracic CT scans:
\begin{itemize}
  \item \textbf{Ground-truth audit and canonical frame verification}:
  \texttt{python -m ribline.scripts.fix\_gt\_annotations} repairs the merged \texttt{R07}/\texttt{R08} annotation in \texttt{RIB\_007}, and \texttt{python -m ribline.scripts.polar\_angle\_stats} computes the lung-apex, spine-midline, and kyphosis-aware polar-angle monotonicity statistics across all 59 annotated scans.
  \item \textbf{Atlas construction and detector training}:
  \texttt{python -m ribline.scripts.build\_atlas}, \texttt{train\_anatomical\_scf}, and \texttt{train\_unet} fit the 24-rib statistical atlas, the 150-tree anatomically anchored SCF, and the 3D U-Net on the 47 training scans defined in \texttt{split.json}.
  \item \textbf{Probability inference and Stage-3 reconstruction}:
  \texttt{python -m ribline.scripts.infer\_anatomical\_scf} and \texttt{infer\_unet} generate voxel-wise probability maps, and \texttt{python -m ribline.scripts.run\_stage3} runs hybrid fusion, skeleton graph extraction, bilateral Viterbi labeling, and \texttt{RibSnake} refinement.
  \item \textbf{Ablation and 5-fold cross-validation}:
  \texttt{python -m ribline.scripts.run\_ablation} reproduces Table~\ref{tab:ablation} on the 12 evaluation scans and the 47 training scans, and \texttt{python -m ribline.scripts.run\_cv5} runs the 59-scan 5-fold cross-validation reported in Section~\ref{sec:cv}.
\end{itemize}

\section{Pretrained Models on Hugging Face}
\label{app:huggingface}

All pretrained checkpoints, deterministic feature configurations, and per-subject evaluation reports are publicly hosted on the Hugging Face Hub and organized in the \texttt{ribline} collection at \url{https://huggingface.co/collections/wq2012/ribline}. Table~\ref{tab:app_hf_models} summarizes the four published model repositories.

\begin{table}[h!]
\vspace{-2mm}
\centering
\caption{Pretrained model repositories published on Hugging Face (\url{https://huggingface.co/collections/wq2012/ribline}). All checkpoints are trained on the 47 SCR training scans; 12-scan and 59-scan 5-fold CV metrics use the symmetric 10\,mm criterion and 50\,mm missed-rib LCD penalty.}
\label{tab:app_hf_models}
\resizebox{\textwidth}{!}{%
\footnotesize
\setlength{\tabcolsep}{3pt}
\begin{tabular}{lp{4.3cm}cccc}
\toprule
\textbf{Hugging Face Repository ID} & \textbf{Model Variant \& Included Checkpoints} & \textbf{Size} & \textbf{12-Scan LCD} & \textbf{12-Scan Acc} & \textbf{5-Fold CV LCD / Acc} \\
\midrule
\texttt{wq2012/ribline-3d-ct-centerline} & Full Hybrid Suite (SCF + 3D U-Net + Pass 1/2 + Atlas) & $529$\,MB & $\mathbf{2.05 \pm 0.48}$\,\textbf{mm} & $\mathbf{97.2\,\%}$ & $\mathbf{3.64 \pm 3.90}$\,\textbf{mm} / $\mathbf{91.8\,\%}$ \\
\texttt{wq2012/ribline-anatomical-scf} & Standalone Anatomically Anchored SCF + Atlas (CPU) & $152$\,MB & $2.88 \pm 1.32$\,mm & $95.1\,\%$ & $5.28 \pm 4.62$\,mm / $86.9\,\%$ \\
\texttt{wq2012/ribline-3d-unet} & Standalone Lightweight 3D U-Net ($340$k params) + Atlas & $1.7$\,MB & $3.40 \pm 1.99$\,mm & $92.0\,\%$ & $7.44 \pm 8.93$\,mm / $78.2\,\%$ \\
\texttt{wq2012/ribline-scf-pass2-autocontext} & Two-Pass Auto-Context SCF Baseline + Atlas & $377$\,MB & $9.36 \pm 10.57$\,mm & $71.0\,\%$ & --- \\
\bottomrule
\end{tabular}%
}
\vspace{-3mm}
\end{table}

\subsection{Repository Descriptions}
\begin{enumerate}
  \item \textbf{\texttt{wq2012/ribline-3d-ct-centerline}} (\url{https://huggingface.co/wq2012/ribline-3d-ct-centerline}):
  The complete flagship suite containing all pretrained models and evaluation artifacts: the anatomically anchored SCF (\texttt{rib\_forest\_anatomical.joblib}, 150 trees, 286 features), the epoch-20 3D U-Net state dictionary (\texttt{rib\_unet\_3d.pth}, $339{,}889$ parameters, \texttt{train\_config\_unet.json}), the Pass-1 and Pass-2 auto-context forests (\texttt{rib\_forest\_pass1.joblib}, \texttt{rib\_forest\_pass2.joblib}), the 24-rib statistical atlas (\texttt{rib\_atlas.joblib}), and the full JSON evaluation logs (\texttt{benchmark\_val12\_fixed\_gt.json}, \texttt{benchmark\_val12\_original\_gt.json}, \texttt{benchmark\_train47.json}, and \texttt{benchmark\_cv5.json}).
  \item \textbf{\texttt{wq2012/ribline-anatomical-scf}} (\url{https://huggingface.co/wq2012/ribline-anatomical-scf}):
  Standalone CPU-only deployment package containing \texttt{rib\_forest\_anatomical.joblib}, \texttt{feature\_config\_anatomical.json}, and \texttt{rib\_atlas.joblib}. Requires no GPU or deep learning runtime at inference time while achieving $2.88 \pm 1.32$\,mm LCD and $95.1\,\%$ labeling accuracy on the 12 evaluation scans ($5.28 \pm 4.62$\,mm and $86.9\,\%$ in 5-fold cross-validation).
  \item \textbf{\texttt{wq2012/ribline-3d-unet}} (\url{https://huggingface.co/wq2012/ribline-3d-unet}):
  Compact deep-learning detector package ($1.7$\,MB total) containing \texttt{rib\_unet\_3d.pth}, \texttt{train\_config\_unet.json}, and \texttt{rib\_atlas.joblib}. Achieves $100.0\,\%$ curve precision (zero false positives) and $3.40 \pm 1.99$\,mm LCD when paired with Stage~3.
  \item \textbf{\texttt{wq2012/ribline-scf-pass2-autocontext}} (\url{https://huggingface.co/wq2012/ribline-scf-pass2-autocontext}):
  Archives the two-pass appearance + auto-context random forest baseline (\texttt{rib\_forest\_pass1.joblib} and \texttt{rib\_forest\_pass2.joblib}) adapted from \cite{wang2013semantic}, illustrating both the limitation of non-anatomical intensity-gated auto-context in thoracic CT and the $+28.0$ percentage-point accuracy gain provided by Stage-3 bilateral Viterbi matching over naive $Z$-sorting.
\end{enumerate}

\subsection{Programmatic Checkpoint Loading}
All checkpoints can be downloaded and instantiated with a single call via \texttt{ribline.hub.load\_pretrained\_models}, or loaded individually via \texttt{huggingface\_hub}:
{\footnotesize
\begin{verbatim}
from ribline import load_pretrained_models

# Download & instantiate atlas, Anatomical SCF, and 3D U-Net
models = load_pretrained_models(
    "wq2012/ribline-3d-ct-centerline", device="cpu"
)
atlas = models["atlas"]
rf_anat, unet = models["forest"], models["unet"]
\end{verbatim}
}

\end{document}